\documentclass[prl,floats,twocolumn,amssymb]{revtex4}

\input epsf
\usepackage{bm}
\usepackage{amsmath}
\usepackage{graphicx}

\begin{document}

\renewcommand{\bottomfraction}{0.7}
\renewcommand{\topfraction}{0.7}
\renewcommand{\textfraction}{0.2}
\renewcommand{\floatpagefraction}{0.7}
\renewcommand{\thesection}{\arabic{section}}

\addtolength{\topmargin}{10pt}


\newcommand{\eq}{\begin{equation}}
\newcommand{\en}{\end{equation}}
\newcommand{\eqa}{\begin{eqnarray}}
\newcommand{\ena}{\end{eqnarray}}
\newcommand{\eqan}{\begin{eqnarray*}}
\newcommand{\enan}{\end{eqnarray*}}
\newcommand{\spz}{\hspace{0.7cm}}
\newcommand{\lbl}{\label}


\def\Bbb{\mathbb}


\newcommand{\Dslash}{{\slash{\kern -0.5em}\partial}}
\newcommand{\Aslash}{{\slash{\kern -0.5em}A}}

\def\sqr#1#2{{\vcenter{\hrule height.#2pt
     \hbox{\vrule width.#2pt height#1pt \kern#1pt
        \vrule width.#2pt}
     \hrule height.#2pt}}}
\def\smallsquare{\mathchoice\sqr34\sqr34\sqr{2.1}3\sqr{1.5}3}
\def\square{\mathchoice\sqr68\sqr68\sqr{4.2}6\sqr{3.0}6}
 
\def\thinspace{\kern .16667em}
\def\punto{\thinspace .\thinspace}
 
\def\xp{x_{{\kern -.2em}_\perp}}
\def\subp{_{{\kern -.2em}_\perp}}
\def\kperp{k\subp}

\def\derpp#1#2{{\partial #1\over\partial #2}}
\def\derp#1{{\partial~\over\partial #1}}

\def\zbar{\overline{z}}
\def\wbar{\overline{w}}
\def\ez{{{\bf e}}_z}
\def\ezbar{{{\bf e}}_{\zbar}}
\def\vF{ v_{_{\rm F}} }
\def\EF{ E_{_{\rm F}} }
\def\nO{ n_{_{\Omega}} }
\def\Kp{\bm{K}_+}
\def\Km{\bm{K}_-}
\def\psiA{\psi_{_{A}}}
\def\psiB{\psi_{_{B}}}
\title{Topological Phases in Graphitic Cones}

\author{Paul~E.~Lammert}\email{lammert@phys.psu.edu} 
\author{Vincent~H.~Crespi}
\affiliation{
     Department of Physics and Center for Materials Physics \\
     The Pennsylvania State University \\
     104 Davey Lab \\
     University Park, Pennsylvania 16802-6300}
\date{August 24, 2000}

\begin{abstract}
The electronic structure of graphitic cones
exhibits distinctive topological features associated with the apical
disclinations.  Aharonov-Bohm magnetoconductance oscillations 
(period $\Phi_0$) are completely absent in
rings fabricated from cones with a single pentagonal disclination.
Close to the apex, the local density of states changes qualitatively, either
developing a cusp which drops to zero at the Fermi energy, or forming a
region of nonzero density across E$_F$, a local metalization of graphene.
\newline
\copyright  2000, The American Physical Society
\end{abstract}
\pacs{73.61.Wp, 73.20.Dx, 73.50.Jt}
%

\maketitle

\begin{figure}
\centerline{\rotatebox{0}{
\resizebox{!}{50 mm}{\includegraphics{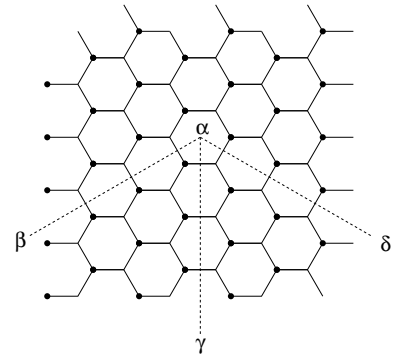}}}}
\caption{
Constructing a cone from a graphene sheet.  Cut out sector 
$\beta\alpha\gamma$ and join $\alpha\beta$ to $\alpha\gamma$ 
to produce a ($\nO=1$) cone with an apical pentagon.  
Excise a second sector and join $\alpha\beta$ to $\alpha\delta$ to 
produce a ($\nO=2$) cone with a smaller opening angle. 
The resulting four-membered ring is very unstable; 
actual cones of that opening angle almost surely contain 
two nearby pentagonal defects instead.  
}
\label{disclination}
\end{figure}

If one or more sectors are excised from a single layer of graphite
(hereafter, {\it graphene}) as illustrated in Figure \ref{disclination},
and the remainder is joined seamlessly, a cone results.  Researchers
discovered how to produce such {\it graphitic cones\/} in 1997\cite{cones}.
We show that the topological nature of the apical defect profoundly
modifies the low-energy electronic structure, with remarkable consequences
for nonlocal transport properties in this new class of nanoscale carbon.  
Rings made from cones exhibit an anomalous Aharonov-Bohm 
effect.  In some cones, the fundamental period is completely absent.
Also, the Fermi-level density of states near the apex is enhanced
and shows distinctive energy dependence.

We classify a cone by its opening
angle, or equivalently the number $\nO$ of sectors removed. All possible
angles have been observed experimentally; we concentrate on $\nO=1$ and
$\nO=2$, which reveal the essential physics.  
Since pentagonal defects are expected to predominate in real cones,
$\nO$ is also typically equal to the number of defects, which must 
be tightly clustered to produce a conical shape.

The gap of a semiconducting tube arises from 
frustration\cite{saito,hamada,dunlap} in the phase of the graphitic 
Fermi-level states as they pass around the tube circumference.  
In this sense, the
semiconducting nanotube is a translational phase defect.  
A cone forms a {\em rotational\/} defect which also frustrates
electronic phase.
In nanotubes, the frustration affects states at the 
two graphene Fermi points in precisely the same way.  
In contrast, the topological phase in a
cone of odd $\nO$ {\it entwines the two Fermi points}.  
Although most of a large cone is just a gently curved graphene 
surface, the global electronic properties are significantly 
disturbed.

Before presenting the new results, we briefly recall the continuum
theory\cite{divincenzo-mele,kane-mele,geo-pert} of graphene, as it forms
the framework for this understanding.  We require a flexible 
real-space continuum formulation to handle structures which are locally 
graphitic but contain non-contractible closed loops, i.e., those 
surrounding the waist of a nanotube or the apex of a cone.  We think of 
the entire cone as having a perfect graphene-like structure locally, 
by considering a pentagonal defect not as a five-membered ring, but as 
a hole in a graphene structure.  In the continuum picture this makes 
sense and the hole can be shrunk to a point.  We can then represent 
defects as pure gauge fields.  
Our continuum description is ideally suited to bringing out 
topological aspects, so long as one exercises care in handling
phenomena on atomic length scales. 

\begin{figure}
\centerline{
\rotatebox{0}{
\resizebox{!}{35 mm}{
\includegraphics{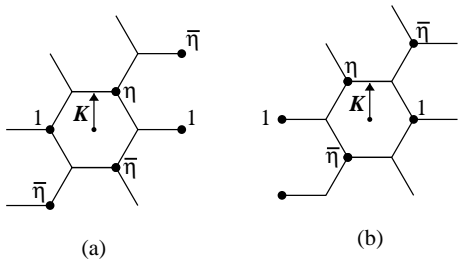}
}}}
\caption{
Tight-binding wavefunctions of graphene Fermi-level states.
Pick a direction
$\bm{K}$ pointing from the center of a hexagon to the middle of one side,
and assign amplitudes as shown, where $\eta = \exp(2\pi i/3)$.  If the
nonzero amplitudes are on the right-hand atoms of bonds cut by $\bm{K}$, as
in (a), the state is $|\bm{K}A\rangle$, if on the left-hand atoms, as in
(b), $|\bm{K}B\rangle$.  There are only two distinct $\bm{K}$.  }
\label{Fermi}
\end{figure}

Graphene's Fermi surface comprises only two points.  Figure 
\ref{Fermi} depicts the Fermi point states in a tight-binding language.
The states are labelled by the direction of a
wave-vector $\bm{K}$ and a sublattice index $A$ or $B$.  The figure
illustrates $|\bm{K} A\rangle$ and $|\bm{K} B\rangle$, where $\bm{K}$
points up the page.  The Fermi level space is four-dimensional, and we
choose $|\bm{K}_\pm A\rangle$ and $|\bm{K}_\pm B\rangle$ as our basis, with
$\bm{K}_- = -\bm{K}_+$.  Under rotation counterclockwise about any hexagon
center by 120$^\circ$, implemented by ${R_6}^2$, these states pick up
simple phases: $ {R_6}^2 |\bm{K} A\rangle = {\eta} |\bm{K} A\rangle$, and
${R_6}^2 |\bm{K} B\rangle = \overline{\eta} |\bm{K} B\rangle$, where $\eta
= \exp(2\pi i/3)$.

We avoid momentum space notions because they are not available in
systems such as cones, which lack translational invariance.  $\bm{K}$ is
only a label for the orientation of the pattern of amplitudes.  However,
whether in a flat graphene sheet, a nanotube, a cone, or any other
graphitic structure, the states near the Fermi level must look like the
four we have constructed, at least locally.  Globally, low-energy states
consist of slowly-varying envelope functions multiplying these four
patterns.  We collect the envelopes into a four-component wavefunction
\begin{equation}
\Psi(x) = 
\begin{pmatrix} \psi_{{_A}} \\ \psi_{{_B}} \end{pmatrix},
\qquad \psi_{_{A, B}}(x) 
= \begin{pmatrix} \psi_{_{A, B +}} \\ \psi_{_{A, B -}} \end{pmatrix},
\end{equation}
where the $\pm$ subscripts refer to $\Kp$ and $\Km$.  This is very similar
to an effective mass description.  We discuss obstructions to a globally
consistent choice of $\Kp$ later.  If that choice is successfully made,
$\Km$, $A$ and $B$ present no further difficulty.

To complete the picture, we need a Hamiltonian.  To avoid complicating
distractions, we first consider simple graphene wherein crystal momentum is
a good quantum number and we can work with just the $\Kp$ components.  In
that case, the Hamiltonian is\cite{divincenzo-mele,kane-mele,geo-pert}
\begin{equation}
H_0 = 
-i\vF (\sigma_1 \partial_x + \sigma_2 \partial_y)
 = -i\vF \bm{\sigma}\cdot\bm{\partial}.
\label{H0}
\end{equation}
The $\sigma$'s are Pauli spin matrices operating in the $A/B$ indices; $A$
is `up,' $B$ is `down,' and the $x$ axis is along $\Kp$.  
For the $\Km$ components of the wavefunction, the appropriate
frame ($x$ and $y$ axes) is rotated through 180$^\circ$
relative to that for $\Kp$.  However, by an 
appropriate change of phases ($\pm \pi/2$), the $\Kp$ frame 
may be used for all components.
The frame need not be locked to the underlying lattice, however, 
a fact which is very useful on nontrivial surfaces such as cones. 
If the coordinate frame $\{ \hat{\bm e}_x,\hat{\bm e}_y \}$ 
in Eq. (\ref{H0}) is rotated by $\theta$ counterclockwise, 
and the wavefunction simultaneously transformed by multiplication by
$\exp{(i\theta \sigma_3/2)}$, the form of the Hamiltonian
is preserved.  This invariance is made local, by introducing 
a gauge field, replacing $\bm{\partial}$ by 
$\bm{\partial} - i\bm{\alpha}$, where 
$\alpha = -\sigma_3 \bm{\partial}\theta/2$.

We imagine taking the sheet in Fig. \ref{disclination}
with wedge $\beta\alpha\gamma$ missing and wrapping it up
to form a cone. The edge $\alpha\gamma$ is then labelled by
azimuthal angle $\phi=0$ and $\alpha\beta$ by $\phi=2\pi$.
What are the appropriate boundary conditions?
Using the four Fermi point patterns described above,
one can see that, up to phases, $|{\Kp}A;\phi=0 \rangle$ is the same
thing as $|{\Km}A;\phi=2\pi \rangle$, and 
$|{\Km}B; \phi=0 \rangle$ the same as $|{\Kp}B; \phi=2\pi \rangle$.
There is no globally coherent distinction between 
$\Kp$ and $\Km$, a remarkable conclusion.
In a sense, just as a M\"obius strip has only one side, 
cones with odd $\nO$ have only one Fermi point!  

Before dealing with the awkward boundary conditions,
notice that the local frames $\{\hat{\bm e}_x,\hat{\bm e}_y\}$
are discontinuous across the seam.  Using the local
rotation invariance mentioned earlier, we orient our frames so that 
$\hat{\bm{e}}_x = \hat{\bm{e}}_\phi$ points tangentially, and 
$\hat{\bm{e}}_y = -\hat{\bm{e}}_r$ toward the apex.
This change introduces a gauge field\cite{strings} (now for general $\nO$)
$$ {\bm \alpha} = -\frac{2\pi \hat{{\bm e}}_\phi}{\ell}
\left( \frac{1}{2} - \frac{\nO}{12}\right) \sigma_3,$$
where $\ell = [1-\nO/6]2\pi r$ is the circumference of the cone at 
distance $r$ from the apex.
The adjustment of the frames makes an additional contribution to the
boundary conditions.  The various ingredients combine to give 
$\Psi(\phi=2\pi) = T(\nO) \Psi(\phi=0)$,
where the operator $T$ (a holonomy) is
\begin{equation}
T(\nO) =  (-1) \exp\left[ 2\pi i \frac{\nO}{4} \tau_2 \right].
\label{holonomy}
\end{equation}
$\tau_2$ is the standard ($2 \leftrightarrow y$) 
Pauli matrix acting on the Fermi point indices $\pm$.
For odd $\nO$, the exponential factor is eliminated by
a singular gauge transformation, 
${\bm \partial} -i{\bm \alpha}
 \mapsto {\bm \partial} -i{\bm \alpha} -i{\bm \beta}$
with 
$${\bm \beta} = \frac{2\pi \hat{{\bm e}}_\phi}{\ell}
\frac{\nO}{4} \tau_2.$$
Recall that in solving the ordinary
Schr\"odinger equation on a flux-threaded ring, 
the vector potential can be eliminated by imposing a 
discontinuous boundary condition 
$\psi(\theta=2\pi) = \exp[ ie \oint \bm{A}\cdot d\bm{l}] \psi(0)$.
Here we perform essentially the reverse procedure, but
our gauge field is proportional to a Pauli matrix.
For even $\nO$, the two Fermi points are not mixed
and the exponential factor is just $-1$. 
So the gauge transformation need not involve $\tau$-matrices,
but merely introduces half a flux quantum of fake magnetic flux.
We deal with the first factor of $-1$ in Eq. (\ref{holonomy}) 
by using anti-periodic boundary conditions. 

With the rotational symmetry more manifest, we can now make a 
partial-wave decomposition of a general spinor as
\begin{equation}
{\Psi}(r,\phi) = \sum_{j} \chi^{(j)}(r) e^{ij\phi},
\end{equation}
and a similar decomposition of the Hamiltonian as 
$H = \sum_j h^{(j)}$.  The total angular momentum
takes on all {\em half-integer\/} values, \hbox{$j = \ldots, -3/2, -1/2,
1/2,\ldots$}.  We work with a finite cone, so that the radial 
wavefunction $\chi^{(j)}$ is in the Hilbert space of 
functions on $(0,R]$ which are square integrable with respect to $r\, dr$,
and the radial Hamiltonian $h^{(j)}$ is of the form
\begin{equation}
h_\nu = \vF 
\begin{pmatrix} 0  & \partial_r + \left(\nu+\frac{1}{2}\right)\frac{1}{r} \\ 
-\partial_r + \left(\nu-\frac{1}{2}\right)\frac{1}{r} & 0 
\end{pmatrix}.
\label{}
\end{equation}
The value of $\nu$ depends on $j$ and the type of 
cone involved:
\begin{equation}
\nu = \begin{cases} j + \frac{\Phi}{\Phi_0} , & \nO = 0; \\
\frac{6}{5} \left( j + \frac{\Phi}{\Phi_0} + 
\frac{\tau_2}{4} \right), & \nO = 1; \\
\frac{3}{2} \left( j- \frac{1}{2} + \frac{\Phi}{\Phi_0} \right), & \nO = 2. \\
\end{cases}
\label{j to nu}
\end{equation}
We include the flat sheet ($\nO=0$, no disclination) as a `control,' and 
also introduce a magnetic flux $\Phi$ through the disclination in order
to study magnetoconductance ($\Phi_0 = h/e$ is the normal flux
quantum).  Notice $\tau_2$ in the $\nO=1$ expression ---
energy eigenstates are superpositions of the two $\bm{K}$.

Comparing the eigenvalue equation for $h_\nu$ to standard recursion
relations for cylinder functions\cite{Bessel,gradshteyn} 
reveals the solutions to be ordinary Bessel functions:
\begin{equation}
\chi_{\nu,1}^\pm(r)
= \begin{pmatrix} J_{\nu-\frac{1}{2}}(kr) \\ 
\pm J_{\nu+\frac{1}{2}}(kr) 
\end{pmatrix}, \quad \nu \ge 0
\label{chi1}
\end{equation}
and
\begin{equation}
\chi_{\nu,2}^\pm(r) = 
\begin{pmatrix} J_{-\nu+\frac{1}{2}}(kr) \\ 
\pm J_{-\nu-\frac{1}{2}}(kr) 
\end{pmatrix}, \quad \nu \le 0.
\label{chi2}
\end{equation}
Thus,
$$
h_\nu  \chi_{\nu,i}^\pm(r) = (\pm \vF k) \chi_{\nu,i}^\pm(r),
\qquad i=1,2.
$$

The indicated restrictions on $\nu$ involve a subtlety.  It
is insufficient simply to ask that $\chi$ and $h_\nu \chi$ be square
integrable.  That requirement says that $\chi_{\nu,1}$ is acceptable for
$-1/2 \leq \nu$ and $\chi_{\nu,2}$ for $\nu \leq 1/2$. 
Also requiring that the radial Hamiltonian $h_\nu$ be self-adjoint,
as it must, forbids both $\chi_{\nu,1}$ and $\chi_{\nu,2}$ in the 
domain of $h_\nu$ at the same time, except for the special case $\nu = 0$.

\begin{figure}
\centerline{
\rotatebox{0}{
\resizebox{85 mm}{!}{
\includegraphics{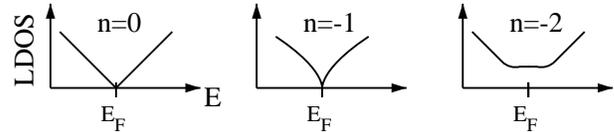}
}}}
\caption{Schematic densities of states for a small 
patch near the apex of a cone, according to Eq. (\ref{LDOS}).
} 
\label{DOSfig}
\end{figure}

Now we turn to observable consequences.
The energy density of states near the apex of a cone shows a remarkable
dependence on the opening angle of the cone.  In fact, the strictly local
density of states diverges as $r\to 0$ for most cones, so we investigate
instead the total density of states on a patch $0 < r \le \delta$ for small
$\delta$.  This quantity is more relevant anyway for comparison to both
experiments and tight-binding computations.  The result is
\begin{equation}
D(E,\delta)  \propto 
\begin{cases}
E \delta^2, &  \nO = 0; \\
E^{3/5} \delta^{8/5}, & \nO = 1; \\ 
\delta, & \nO = 2;
\end{cases}
\label{LDOS}
\end{equation}
as illustrated schematically in figure \ref{DOSfig}.

Eq. (\ref{LDOS}) is derived as follows.  From the large-$x$ 
asymptotic formula 
$J_n(x) \sim \left({2}/{\pi x}\right)^{1/2} \cos\left[ x -
\left(n + \frac{1}{2} \right) \frac{\pi}{2} \right]$,
we deduce a normalization factor of $ c = \left( {\pi
k}/{2R} \right)^{1/2} $ for our $\chi_{\nu,i}^\pm$ functions.  The only
values of $j$ which contribute significantly to the density of states near
$r=0$ are $j=\pm 1/2$ for $\nO = 0,1$ and $j=1/2$ for $\nO=2$, and the
leading order behavior of $J_n$ is used to integrate over a small disk.
Finally, the spacing of $k$ values is $\Delta k = \pi/R$ ($2\pi/R$
for $j=1/2$ and $\nO = 2$).  Thus, the total density of states in the
$\delta$-disk contributed by states associated with Bessel function order
$n$ is
$$
D(k,\delta,n) \sim c^2 \frac{|k \delta|^{2 n} \delta^2}{\vF \Delta k}
\propto  |E|^{2 n + 1} \delta^{2(n+1)}.
$$
Inserting appropriate values for $n$, we get Eq. (\ref{LDOS}).

Strikingly, the low-energy density of states for the $\nO=1$ cone has a
cusp at $E_F$, and the cone with two sectors missing has a non-zero density
of states near the Fermi level.  An actual cone with the $\nO=2$ opening
angle probably actually has a pair of $\nO=1$ defects.  However, if the two
pentagons are very near each other, we expect this conclusion to remain
valid. These remarkable results should be observable in STM spectroscopy,
so long as they are distinguished from the local effect of bond strain in
the pentagonal defect(s).

Although the continuum description lacks atomistic detail, the
qualitative conclusions agree well with an earlier atomistic 
computation\cite{japanese} of (apical) local densities of states 
using a one-orbital tight-binding model for similar structures.  
Direct comparison with our Eq. (\ref{LDOS}) can be made, bearing 
in mind the low-energy restriction on our calculations.  In references
\onlinecite{japanese}, it was suggested that the states contributing to the
non-zero density of states at $\EF$ for $\nO=2$ are (power-law) localized.
Our calculations show that they are extended states which are 
enhanced in the vicinity of the apex.

A mesoscopic normal metal ring in a perpendicular magnetic field exhibits
oscillations as the flux through the ring is varied\cite{Aronov-Sharvin}.
The longest period of oscillation is one normal flux quantum $\Phi_0 =
h/e$, corresponding essentially to the Aharonov-Bohm effect, and first
observed in this context in the mid-1980's\cite{webb}. 
Manufacturing such a ring from a graphitic cone seems possible.  
The apex could be etched away with acid, or cut off with an STM.
Such manipulation does not change $\nO$ because it measures a
topological property --- the number of missing sectors, or 
equivalently, the opening angle.

The magnetoconductance of such conical rings also shows profound sensitivity
to the opening angle.  
Although we could argue directly from the holonomies, it is perhaps simpler
to appeal to Eq. (\ref{j to nu}).  These expressions are still valid
for a ring, even though the subsequent analysis was specific to a cone with
only the apical point removed.  For $\nO= 0$ and $\nO=2$, the Hamiltonian
is $SU(2)_K$ invariant (the $SU(2)$ group acts on the ${\bm{K}_\pm}$ indices),
so that the two Fermi points produce two
independent branches of excitations which respond identically to a magnetic
flux.  Each branch exhibits its own magnetoconductance, which has period
$\Phi_0$, just as in an ordinary metal, and the phases are identical.  The
total response therefore also has fundamental period $\Phi_0$.

In an $\nO=1$ ring, on the other hand, {\it the $\Phi_0$ component of the
oscillations is extinguished.}  From Eq. (\ref{j to nu}), we see
that the $\tau_2 = +1$ ($\tau_2 = -1$) branch behaves as though it were
subjected to a flux of $\Phi = \Phi_0/4$ ($\Phi = -\Phi_0/4$).  This
relative shift of the magnetoconductance curves causes cancellation
of the $\Phi_0$ periodic component. Cones with intact apices
will also show novel magnetic phenomena, including the prospect for
field-tuneable radial charge density waves.

Our neglect of disorder and inelastic scattering in this discussion
should be permissible up to the micron length scale.
The elastic scattering length, $\ell_p$, in single-wall 
nanotubes\cite{ballistic} is believed to range up to many
microns and there is direct evidence that the phase coherence length
($\ell_\phi$) at room temperature is also that long\cite{coherent}.  A
graphitic ring would be expected to have a somewhat reduced $\ell_p$, due
to rough edges, but $\ell_\phi$ is likely comparable to that in the tube.

In conclusion, we demonstrate that graphene cones, which have been
experimentally produced, but not yet adequately studied, comprise a new
class of nanoscale carbon wherein phase
frustration induces profound modulations of the low-energy electronic
properties. These rotational phase defects (induced by topological lattice
defects) leave distinctive and non-trivial local and global imprints on the
electronic structure of graphitic cones.  We predict two specific phenomena
flowing from this observation: apical enhancement of density of states and
an anomalous magnetoconductance.  

We gratefully acknowledge the David and Lucile Packard Foundation and the
National Science Foundation through grant DMR--9876232.


\begin{thebibliography}{}


\bibitem{cones} A.~Krishnan, E.~Dujardin, M.~M.~J.~Treacy,
J.~Hugdahl, S.~Lynum, and T.~W.~Ebbesen, Nature {\bf 388}, 451 (1997).

\bibitem{saito} R.~Saito, M.~Fujita, G.~Dresselhaus, and
M.~S.~Dresselhaus, Appl. Phys. Lett. {\bf 60}, 2204 (1992).

\bibitem{hamada} 
N.~Hamada, S.~Sawada, and A.~Oshiyama, Phys. Rev. Lett.
{\bf 68}, 1579 (1992).

\bibitem{dunlap} B. I. Dunlap and C. T. White, Phys. Rev. Lett. 
{\bf 68}, 631 (1992).

\bibitem{divincenzo-mele} 
D.~P.~DiVincenzo and E.~J.~Mele, Phys. Rev. B {\bf 29}, 1685 (1983).

\bibitem{kane-mele} C.~L.~Kane and E.~J.~Mele, Phy. Rev. Lett. {\bf 78},
1932 (1997).

\bibitem{geo-pert} P.~E.~Lammert and V.~H.~Crespi, Phys. Rev. B
{\bf 61}, 7308 (2000).  (Note differing $A/B$ convention.)

\bibitem{strings}
In another context, ${\bm \alpha}$, the {\it spin connection\/},
expresses the influence of gravity on a spinor field.
We have a simple model of a massless Fermion propagating
near a (very!) massive cosmic string in $2+1$ dimensions
[see, e.g., A.~Vilenkin and E.~P.~S.~Shellard, 
{\it Cosmic Strings and Other Topological Defects} 
(Cambridge University Press, Cambridge, 1994)].
The string produces a curvature singularity.

\bibitem{Bessel} See, e.g.,
G.~Arfken, {\it Mathematical Methods for Physicists}, 
3rd Ed. (Academic Press, Orlando 1985);
N.~Temme, {\it Special Functions, An Introduction
to the Classical Functions of Mathematical Physics}
(Wiley, New York 1996).

\bibitem{gradshteyn} 
I.~S.~Gradshteyn and I.~M.~Rhyzik,
{\it Table of Integrals, Series, and Products}, corrected
and enlarged ed. (Academic Press, San Diego 1980), Eq. 8.471.

\bibitem{japanese} R.~Tamura and M.~Tsukada, 
Phys. Rev. B {\bf 52}, 6015 (1995),
Phys. Rev. B {\bf 49}, 7697 (1994); 
R.~Tamura, K.~Akagi, and M.~Tsukada, 
Phys. Rev. B {\bf 56}, 1404 (1997).

\bibitem{Aronov-Sharvin} A.~G.~Aronov, and Yu.~V.~Sharvin,
Rev. Mod. Phys. {\bf 59}, 755 (1987).

\bibitem{webb} R.~A.~Webb, S.~Washburn, C.~P.~Umbach, and
R.~B.~Laibowitz, Phys. Rev. Lett. {\bf 54}, 2696 (1985).

\bibitem{ballistic}
C.~T.~White and T.~N.~Todorov, Nature {\bf 393}, 240 (1998).

\bibitem{coherent}
S.~J.~Tans, M.~H.~Devoret, H.~Dai, A.~Thess, R.~E.~Smalley,
L.~J.~Geerligs, and C.~Dekker, Nature {\bf 386}, 474 (1997).

\end{thebibliography}
\end{document}